\documentclass[showpacs,preprintnumbers,amsmath,amssymb,
superscriptaddress,floatfix]{revtex4}
\usepackage{graphicx}
\usepackage{amsmath}
\usepackage{bm}
\usepackage{mathrsfs}
\usepackage{color}
\usepackage{slashed}
\usepackage{dcolumn}

\newcommand{\be}{\begin{equation}}
\newcommand{\ee}{\end{equation}}
\newcommand{\ba}{\begin{eqnarray}}
\newcommand{\ea}{\end{eqnarray}}
\newcommand{\nn}{\nonumber}

\newcommand{\la}{\langle}
\newcommand{\ra}{\rangle}

\begin{document}

\title{The pseudoscalar meson and baryon octet interaction
with strangeness $S=-2$ in the unitary coupled-channel
approximation}

\author{Bao-Xi Sun}
\email{sunbx@bjut.edu.cn}
\affiliation{Faculty of Science, Beijing University
of Technology, Beijing 100124, China}

\author{Xin-Yu Liu}
\affiliation{Faculty of Science, Beijing University
of Technology, Beijing 100124, China}


\begin{abstract}
The interaction of the pseudoscalar meson and the baryon octet is
investigated by solving the Bethe-Salpeter equation in the infinite and finite volume respectively. It is found that there is a resonance state generated dynamically, which owns a mass about 1550MeV and a large decay width of 120-200MeV. This resonance state couples strongly to the $\pi \Xi$ channel. Therefore, it might not correspond to the $\Xi(1620)$ particle announced by Belle collaboration. At the same time, this problem is studied in the finite volume, and an energy level at 1570MeV is obtained, which is between the $\pi \Xi$ and $\bar{K}\Lambda$ thresholds and independent of the cubic box size.
\end{abstract}


\maketitle

\section{Introduction}

The experimental data on double strange baryons $\Xi(1620)$ and $\Xi(1690)$ are scarce, and the spin-parity of them are not determined, so they are labeled with one star and three stars respectively in the review of Particle Data Group\cite{PDG}.
Recently, the $\Xi(1620)$ particle has been reported to be observed in the decay of $\Xi_{c}^+ \rightarrow \Xi^{-}\pi^+ \pi^+$ by Belle collaboration, and the mass and decay width are measured as
\ba
M&=&1610.4\pm6.0(stat){}^{+6.1}_{-4.2}(syst)MeV, \nn \\
\Gamma&=&59.9\pm4.8{stat}{}^{+2.8}_{-7.1}(syst)MeV,
\ea
respectively. Moreover, there are also some evidences of the $\Xi(1690)$ particle with the same data sample\cite{Belle}.

The masses of these two particle are about 300MeV higher than that of the $\Xi$ hyperon, and thus they can be regarded as excited states of the $\Xi$ hyperon. However they are
difficult to be described within the framework of the constituent
quark model.

In Ref.~\cite{Ramos:2002xh}, the $\Xi(1620)$ particle was assumed to be a resonance state of the pseudoscalar meson and baryon octet with strangeness $S=-2$ and spin $J=1/2$ in the unitary coupled-channel approximation of the Bethe-Salpeter equation. It shows that this resonance state couples strongly to the $\pi \Xi$ and $\bar{K}\Lambda$ channels, and its width is sensitive to the subtraction constants related to these two channels.
The properties of $\Xi(1620)$ was studied by solving the Bethe-Salpeter equation in Ref.~\cite{Guoxinheng}, Apparently, the method is the same as that used in Ref.~\cite{Ramos:2002xh}, but with the kernel introduced by the vector meson exchange
interaction.
It manifests that the $\Xi(1620)$ particle might be a $\bar{K}\Lambda$ or $\bar{K}\Sigma$ bound state.
Moreover, the decay width of $\Xi(1620)\rightarrow \Xi \pi$ is calculated, where $\bar{K}\Lambda$ and $\bar{K}\Sigma$ are treated as the intermediate state respectively, and the results indicates that the component of $\bar{K}\Lambda$ is larger than $\bar{K}\Sigma$ in the $\Xi(1620)$ particle.
Sequently, the radiative decay process of $\Xi(1620)$ is analyzed systematically in Ref.~\cite{Huang:2021ahp} by assuming the $\Xi(1620)$ particle to be a $\bar{K}\Lambda$ and $\bar{K}\Sigma$ bound state with spin-parity $J^P=1/2^-$.
In Ref.~\cite{Hyodo},  a series of non-leptonic weak decays of $\Xi_c$ into $\pi^+$ and a meson-baryon final state are discussed, and the invariant mass distribution of the meson-baryon final state is analyzed within three different chiral scheme. However, it is found that the peak appeared in the $\pi \Xi$ and $\bar{K} \Lambda$ spectra is more possible to be $\Xi(1690)$, but not the $\Xi(1620)$ particle.
The unitary coupled-channel approximation of Bethe-Salpeter equation in the finite volume has made a great success in the study of the meson-meson interaction\cite{Doringmeson,AlbertoKD,Albaladejopipi,Gengf1285} and the meson-baryon interaction\cite{AlbertoKN}. Actually, a scheme to simulate the Lattice data in order to obtain the kernel of Bethe-Salpeter equation in the unitary coupled-channel approximation is proposed in these articles, which is called {\sl the inverse problem}. However, the parameters in the kernel are fit to the calculation results of the Bethe-Salpeter equation in the finite volume, but not the {\sl real} Lattice data.
Finally, an attempt has been made to fit the lattice finite volume energy levels from $\pi \eta$ scattering and the properties of $a_0(980)$ is evaluated\cite{Guozhpieta}. This method is also extended to study the interaction of the $\pi D$, $\eta D$ and $\bar{K} D_S$ channels in $J^P=0^{+}$ in the finite volume by fitting the lattice QCD calculation results\cite{Albaladejo2400}.

In this work, the interaction between the pseudoscalar meson and baryon octet with strangeness $S=-2$ will be studied, and then the Bethe-Salpeter equation in the infinite and finite volume will be solved within the unitary coupled-channel approximation respectively. We will try to distinguish whether there are resonance state generated dynamically or not, and if so, whether the resonance state can be treated as a counterpart of the $\Xi(1620)$ particle.

This article is organized as follows. In
Section~\ref{sect:framework}, the potential of the pseudoscalar
meson and baryon octet is constructed.
In Section~\ref{sect:BS}, a basic formula on how to solve the
Bethe-Salpeter equation in the unitary coupled-channel approximation is shown. Sequently, the pole position in the complex energy plane is obtained by solving the Bethe-Salpeter equation in the infinite volume, and its coupling constants to different channels are calculated.
The chiral unitary approach in a finite box is introduced in Section \ref{sect:finite}
Finally, the summary is given
in Section~\ref{sect:summary}.

\section{Framework}
\label{sect:framework}

The effective Lagrangian of the pseudoscalar meson and the baryon
octet can be written as
\ba \label{eq:piNLargrangian} {\sl L}=\la \bar{B}(i\gamma_\mu D^\mu
- M) B \ra+\frac{D/F}{2}  \la \bar{B} \gamma_\mu \gamma_5
[u^\mu,B]_{\pm}     \ra. \ea
In the above equation, the symbol $\langle...\rangle$ denotes the
trace of matrices in the $SU(3)$ flavor space, and $D^\mu B =
\partial^\mu B + \frac{1}{2} \left[ [u^\dagger, \partial^\mu u], B
\right]$ with $u^2=U=\exp \left( i\frac{\Phi}{f_0} \right)$ and
$u^\mu = iu^\dagger \partial^\mu u - i u \partial^\mu u^\dagger$,
where $f_0$ is the meson decay constant in the chiral limit.

The matrices of the pseudoscalar meson and the baryon octet are
given as follows

\begin{align}
\Phi={}&\sqrt{2}
\begin{pmatrix}
\frac{1}{\sqrt{2}}\pi^{0}+\frac{1}{\sqrt{6}}\eta & \pi^{+} & K^{+}\\
\pi^{-} &  -\frac{1}{\sqrt{2}}\pi^{0}+\frac{1}{\sqrt{6}}\eta & K^{0}\\
K^{-} & \bar{K}^{0} & -\frac{2}{\sqrt{6}}\eta
\end{pmatrix}
 \label{eq:mesons matrices}
\end{align}
and
\begin{align}
B={}&
\begin{pmatrix}
\frac{1}{\sqrt{2}}\Sigma^{0}+\frac{1}{\sqrt{6}}\Lambda & \Sigma^{+} & p\\
\Sigma^{-} &  -\frac{1}{\sqrt{2}}\Sigma^{0}+\frac{1}{\sqrt{6}}\Lambda & n\\
\Xi^{-} & \Xi^{0} & -\frac{2}{\sqrt{6}}\Lambda
\end{pmatrix}.
 \label{eq:baryons matrices}
\end{align}

\begin{figure}[htb]
\begin{center}
\includegraphics[width=0.65\textwidth]{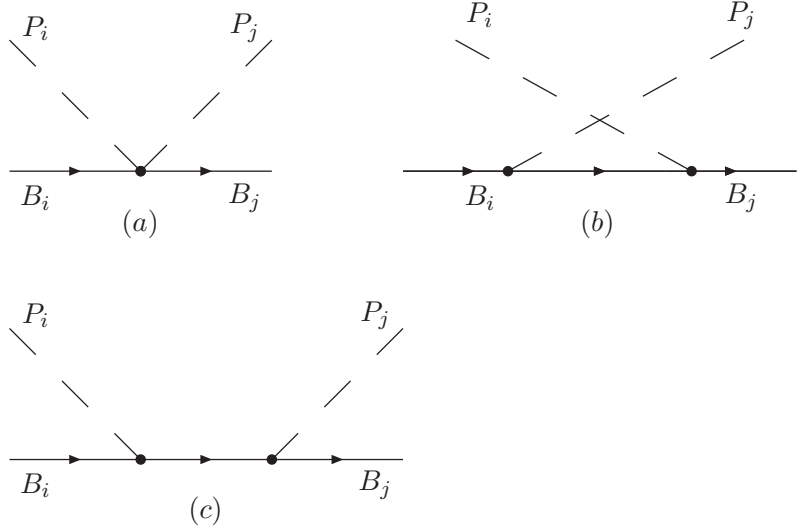}
\end{center}
\caption{Feynman diagrams of the pseudoscalar meson-baryon octet
interaction. $(a)$ contact term, $(b)$ $u-$ channel and $(c)$ $s-$
channel.} \label{fig:Feynman}
\end{figure}

The first term in the Lagrangian in Eq.~(\ref{eq:piNLargrangian})
supplies the contact interaction of the pseudoscalar meson and the
baryon octet, which is usually called as Weinberg-Tomozawa term,
while the other terms which are relevant to the coefficients
$D$ and $F$ give a contribution to the $s-$ and $u-$ channel
interactions, as shown in Fig.~\ref{fig:Feynman}.

The Weinberg-Tomozawa contact term of the pseudoscalar
meson and the baryon octet takes the form of
\begin{align}
\label{eq:wtcontact}
V^{con}_{ij}={}&-C_{ij}\frac{1}{4f_i f_j}(2\sqrt{s}-M_{i}-M_{j})
\left(\frac{M_{i}+E}{2M_{i}}\right)^{\frac{1}{2}}
\left(\frac{M_{j}+E^{\prime}}{2M_{j}}\right)^{\frac{1}{2}},
\end{align}
where $\sqrt{s}$ is the total energy of the system, $M_{i}$ and
$M_{j}$ denote the initial and final baryon masses, $E$ and $E^\prime$ stand for the initial and final baryon
energies in the center of mass frame, respectively.
The coefficient $C_{ij}$ for the sector of strangeness $S=-2$ and
charge zero is listed in Table~\ref{table:coef_S0Q0}, Moreover, we
assume the values of the decay constants are only relevant to the
pseudoscalar meson with $f_{\eta}=1.3f_{\pi}$, $f_K=1.22 f_{\pi}$
and $f_{\pi}=92.4$MeV, as given in Refs.~\cite{Inoue,Bruns,DongSun,SunZhao}.

In the interaction of pseudoscalar meson and baryon octet, the contact potential originated from Weinberg-Tomozawa term plays a dominant role, and the correction from the $s-$ and $u-$ channel potentials can be neglected.
The Weinberg-Tomozawa term of the
pseudoscalar meson and the baryon octet are only related to the
Mandelstam variable $s$, therefore, it only gives a contribution to
the S-wave amplitude in the scattering process of the pseudoscalar
meson and the baryon octet.

\begin{table}[htbp]
\begin{tabular}{c|cccccc}
\hline
 $C_{ij}$          &$\pi^+ \Xi^-$ & $\pi^0 \Xi^0$ & $\bar{K}^0 \Lambda$ & $K^- \Sigma^+$ & $\bar{K}^0 \Sigma^0$ & $\eta \Xi^0$    \\
\hline
 $\pi^+ \Xi^-$      &$1$ & $-\sqrt{2}$ & $-\sqrt{\frac{3}{2}}$ & $0$  & $-\frac{1}{\sqrt{2}}$ & $0$  \\
 $\pi^0 \Xi^0$      &$$  & $0$ & $\frac{\sqrt{3}}{2}$ & $-\frac{1}{\sqrt{2}}$  & $-\frac{1}{2}$ & $0$  \\
$\bar{K}^0 \Lambda$        &$$ & $$ & $0$ & $0$  & $0$ & $-\frac{3}{2}$  \\
$K^- \Sigma^+$            &$$ & $$ & $$ & $1$  & $-\sqrt{2}$ & $-\sqrt{\frac{3}{2}}$  \\
$\bar{K}^0 \Sigma^0$            &$$ & $$ & $$ & $$  & $0$ & $\frac{\sqrt{3}}{2}$    \\
$\eta \Xi^0$             &$$ & $$ & $$ & $$  & $$ & $0$    \\
 \hline \hline
\end{tabular}
\caption{The coefficients $C_{ij}$ in the pseudoscalar meson and
baryon octet interaction with strangeness $S=-2$ and charge $Q=0$,
$C_{ji}=C_{ij}$.} \label{table:coef_S0Q0}
\end{table}

In the sector of strangeness $S=-2$ and isospin $I=\frac{1}{2}$, the
wave function in the isospin space can be written as
\be
| \pi \Xi; \frac{1}{2}, \frac{1}{2} \ra = \sqrt{\frac{2}{3}} |\pi^+
\Xi^- \ra - \sqrt{\frac{1}{3}} | \pi^0 \Xi^0 \ra,
 \ee

\be | \eta \Xi; \frac{1}{2}, \frac{1}{2}\ra = |\eta \Xi^0 \ra,
 \ee

\be |\bar{K} \Lambda; \frac{1}{2}, \frac{1}{2} \ra =|\bar{K}^0 \Lambda \ra, \ee
and
 \be |\bar{K} \Sigma; \frac{1}{2}, \frac{1}{2} \ra =
-\sqrt{\frac{2}{3}} |K^- \Sigma^+ \ra + \sqrt{\frac{1}{3}} |\bar{K}^0
\Sigma^0 \ra, \ee respectively. Thus the coefficients $C_{ij}$ in
the Weinberg-Tomozawa contact potential of the pseudoscalar meson
and the baryon octet can be obtained in the isospin space, which are
summarized in Table~\ref{table:coef_Sm2I12}.

\begin{table}[htbp]
\begin{tabular}{c|cccc}
\hline
 $C_{ij}$          &$\pi \Xi$ &  $\bar{K} \Lambda$ & $\bar{K} \Sigma$ & $\eta \Xi$    \\
\hline
 $\pi \Xi$      & $2$ & $-\frac{3}{2}$  & $-\frac{1}{2}$ & $0$  \\
$\bar{K} \Lambda$        &$$ &  $0$  & $0$ & $-\frac{3}{2}$  \\
$\bar{K} \Sigma$             & $$ & $$  & $2$ & $\frac{3}{2}$    \\
$\eta \Xi$              & $$ & $$  & $$ & $0$    \\
 \hline \hline
\end{tabular}
\caption{The coefficients $C_{ij}$ in the pseudoscalar meson and
baryon octet interaction with strangeness $S=-2$ and isospin $I=1/2$,
$C_{ji}=C_{ij}$.} \label{table:coef_Sm2I12}
\end{table}

In the case of isospin $I=3/2$ and strangeness $S=-2$, the interaction of the pseudoscalar meson and the baryon octet are repulsive,
therefore, no resonance states can be generated dynamically.

\section{Bethe-Salpeter equation}
\label{sect:BS}

The Bethe-Salpeter equation can be expanded as
\ba
\label{eq:1903021709}
T &=&V+VGT \nn \\
 &=&V+VGV+VGVGV+....
\ea
When the Bethe-Salpeter equation in Eq.~(\ref{eq:1903021709}) is
solved, only the on-shell part of the potential $V_{ij}$ gives a contribution to the amplitude of
the pseudoscalar meson and the baryon octet, and the off-shell part
of the potential can be reabsorbed by a suitable renormalization of
the decay constants of mesons $f_i$ and $f_j$.
More detailed discussion can be found in
Refs.~\cite{Oller97,Ramos97,DongSun,SunZhao}.
If the relativistic kinetic correction of the loop function of the
pseudoscalar meson and the baryon octet is taken into account, the
loop function $G_{l}$ can be written as
\begin{align}
G^D_{l}={}&i\int\frac{d^{d}q}{(2\pi)^{4}}\frac{\rlap{/}q+M_{l}}
{q^{2}-M_{l}^{2}+i\epsilon}\frac{1}{(P-q)^{2}-m_{l}^{2}+i\epsilon},
 \label{eq:G}
\end{align}
with $P$ the total momentum of the system, $m_{l}$ the meson mass,
and $M_{l}$ the baryon mass, respectively.

The loop function in Eq.~(\ref{eq:G}) can be calculated in the
dimensional regularization (See Appendix 1 of Ref.~\cite{DongSun}
for details), and thus the loop function takes the form of
\begin{equation}
\begin{aligned}
G^D_{l}={}&\frac{\gamma_{\mu}
P^{\mu}}{32P^{2}\pi^{2}}\left[(a_{l}+1)(m_{l}^{2}-M_{l}^{2})+(m_{l}^{2}\ln\frac{m_{l}^{2}}{\mu^{2}}-M_{l}^{2}\ln\frac{M_{l}^{2}}{\mu^{2}})\right]\\&
+\left(\frac{\gamma_{\mu}
P^{\mu}[P^{2}+M_{l}^{2}-m_{l}^{2}]}{4P^{2}M_{l}}+\frac{1}{2}\right)G_{l}^{\prime},
\end{aligned} \label{eq:Our G}
\end{equation}
where $a_l$ is the subtraction constant and $\mu$ is the
regularization scale, and $G_{l}^{\prime}$ is the loop function in
Ref.~\cite{Oller},
\begin{eqnarray}
G^\prime_{l}(s) &=& \frac{2 M_l}{16 \pi^2} \left\{ a_l(\mu) + \ln
\frac{m_l^2}{\mu^2} + \frac{M_l^2-m_l^2 + s}{2s} \ln
\frac{M_l^2}{m_l^2} + \right. \nonumber \\ & &  \phantom{\frac{2
M}{16 \pi^2}} + \frac{\bar{q}_l}{\sqrt{s}} \left[
\ln(s-(M_l^2-m_l^2)+2\bar{q}_l\sqrt{s})+
\ln(s+(M_l^2-m_l^2)+2\bar{q}_l\sqrt{s}) \right. \nonumber  \\
& & \left. \phantom{\frac{2 M}{16 \pi^2} +
\frac{\bar{q}_l}{\sqrt{s}}} \left. \hspace*{-0.3cm}-
\ln(-s+(M_l^2-m_l^2)+2\bar{q}_l\sqrt{s})-
\ln(-s-(M_l^2-m_l^2)+2\bar{q}_l\sqrt{s}) \right] \right\},
\label{eq:gpropdr}
\end{eqnarray}
with $\bar{q}_l$ the three-momentum of the meson or the baryon in
the center of mass frame.

Since the total three-momentum $\vec{P}=0$ in the center of mass
frame, only the $\gamma_{0} P^{0}$ part remains in Eq.~(\ref{eq:Our
G}).
Similarly, This matrix $\gamma_{0}$ can be replaced by the unit
matrix $I$ since the $U(p_i,\lambda_i)$ and $\bar{U}(p_j,\lambda_j)$
denote the wave functions of the initial and final baryons,
respectively. Thus the loop function of the intermediate
pseudoscalar meson and baryon octet becomes
\begin{equation}
\begin{aligned}
G^D_{l}={}&\frac{\sqrt{s}}{32\pi^{2}s}\left[(a_l+1)(m_{l}^{2}-M_{l}^{2})+(m_{l}^{2}ln\frac{m_{l}^{2}}{\mu^{2}}-M_{l}^{2}ln\frac{M_{l}^{2}}{\mu^{2}})\right]\\&
+\left(\frac{s+M_{l}^{2}-m_{l}^{2}}{4M_{l}\sqrt{s}}+\frac{1}{2}\right)G_{l}^{\prime}.
\end{aligned}
\label{eq:Our G result}
\end{equation}

The off-shell part of the potential is reabsorbed in a
renormalization process, so the decay constants of mesons, the masses of
intermediate baryons all take physical values when the
Bethe-Salpeter equation is solved.

In the calculation of the present work, we make a transition of
\ba \tilde{V}&=&V~\sqrt{M_i M_j}, \nn \\
    \tilde{G}_l&=&G_l/M_l,
\ea
so the scattering amplitude
\be \tilde{T}=[1-\tilde{V}\tilde{G}]^{-1}\tilde{V} \ee becomes
dimensionless.

The subtraction constants in the loop function of Eqs.~(\ref{eq:gpropdr}) and (\ref{eq:Our G result}) are listed in Table~\ref{table:subtraction}, which are the same as those used in Ref.~\cite{Ramos:2002xh}.
With these subtraction constants and the regularization scale $\mu=630$MeV, the amplitudes of pseudoscalar meson and baryon octet are evaluated by solving the Bethe-Salpeter equation in the unitary coupled-channel approximation.
A pole is detected around 1550MeV in the complex energy plane and the pole position and coupling constants obtained with the loop functions in Eqs.~(\ref{eq:gpropdr}) and (\ref{eq:Our G result}) are summarized in Tables~\ref{table:osetpoleposition} and \ref{table:ourpoleposition}, respectively.
Since the real part of the pole position is higher than the $\pi \Xi$ threshold, and lower than the $\bar{K} \Lambda$ threshold, it lies in the second Riemann sheet and can be regarded as a resonance state with strangeness $S=-2$ and isospin $I=1/2$.
When the values of these subtraction constants change, the mass of this resonance state changes slightly, while the decay width of it changes in the range of 120-200MeV. Apparently, both the mass and the decay width of this resonance state are far away from the experimental value supplied by Belle collaboration, and the results are also different from those in Ref.~\cite{Ramos:2002xh}.

The coupling constants are calculated according to
\be
\frac{g_i g_j \sqrt{M_i M_j}} {\sqrt{s}-\sqrt{s_0}}=\tilde{T},
\ee
with $\sqrt{s_0}$ the pole position in the complex energy plane.
This resonance state couples strongly to the $\pi \Xi$ channel, as listed in
Tables~\ref{table:osetpoleposition} and \ref{table:ourpoleposition}.

\begin{table}[htbp]
\begin{tabular}{c|cccc}
\hline \hline
            & $a_{\pi\Xi}$ & $a_{\bar{K}\Lambda}$  & $a_{\bar{K}\Sigma}$ & $a_{\eta\Xi}$  \\
\hline
 Set 1      & -2.0 & -2.0  & -2.0 & -2.0  \\
 Set 2      & -2.2 & -2.0  & -2.0 & -2.0  \\
 Set 3      & -2.0 & -2.2  & -2.0 & -2.0  \\
 Set 4      & -2.5 & -1.6  & -2.0 & -2.0  \\
 Set 5      & -3.1 & -1.0  & -2.0 & -2.0  \\
 \hline \hline
\end{tabular}
\caption{The subtraction constants $a_{ij}$ used in the calculation, where the regularization scale takes the value of 630MeV, i.e., $\mu=630$MeV.} \label{table:subtraction}
\end{table}

\begin{table}[htbp]
\begin{tabular}{c|c|c|c|c|c}
\hline \hline
                       & Set 1     & Set 2  & Set 3 & Set 4 & Set 5  \\
\hline
 Pole position         & 1566-i119 & 1557-i99  & 1558-i113 & 1558-i83 & 1553-i60  \\

\hline
 $g_{\pi\Xi}$          & 2.2-i1.5    & 2.2-i1.3   & 2.1-i1.5 & 2.2-i2.4  & 2.1-i0.8  \\
 $g_{\bar{K}\Lambda}$  & -1.8+i0.6   & -1.8+i0.5  & -1.7+i0.6 & -1.9+i0.5 & -2.1+i0.4  \\
 $g_{\bar{K}\Sigma}$   & -0.5+i0.3   & -0.5+i0.3  & -0.5+i0.3 & -0.6+i0.3  & -0.7+i0.2 \\
 $g_{\eta\Xi}$         & 0.1-i0.3     & 0.1-i0.3     & 0.2-i0.3 & -0.0-i0.1 & -0.3-i0  \\
 \hline \hline
\end{tabular}
\caption{The pole position and corresponding coupling constants $g_{i}$ for different parameter sets calculated with the loop function in Eq.~(\ref{eq:gpropdr}), where the pole position in the complex energy plane is in units of MeV.} \label{table:osetpoleposition}
\end{table}

\begin{table}[htbp]
\begin{tabular}{c|c|c|c|c|c}
\hline \hline
                       & Set 1     & Set 2  & Set 3 & Set 4 & Set 5  \\
\hline
 Pole position         & 1557-i104 & 1550-i89  & 1552-i100 & 1551-i78 & 1546-i60  \\

\hline
 $g_{\pi\Xi}$          & 2.2-i1.4    & 2.2-i1.2   & 2.1-i1.4 & 2.2-i1.0  & 2.1-i1.0  \\
 $g_{\bar{K}\Lambda}$  & -1.7+i0.5   & -1.7+i0.5  & -1.7+i0.5 & -1.8+i0.5 & -2.0+i0.4  \\
 $g_{\bar{K}\Sigma}$   & -0.5+i0.3   & -0.5+i0.3  & -0.5+i0.3 & -0.6+i0.3  & -0.6+i0.2 \\
 $g_{\eta\Xi}$         & 0.1-i0.3     & 0.1-i0.2     & 0.2-i0.3 & 0.0-i0.1 & -0.2-i0.0  \\
 \hline \hline
\end{tabular}
\caption{The pole position and corresponding coupling constants $g_{i}$ for different parameter sets calculated with the loop function in Eq.~(\ref{eq:Our G result}), where the pole position in the complex energy plane is in units of MeV.} \label{table:ourpoleposition}
\end{table}

\section{The chiral unitary approach in a finite box}
\label{sect:finite}

In order to obtain the energy level in a finite box, the loop function in Eq.~(\ref{eq:Our G result}) should be be replaced by a $\bar{G}$ when the Bethe-Salpeter equation is solved, where
\begin{align}
\bar{G}(E)=G^D(E)+\lim_{q_{\rm max}\to \infty}\Bigg[\frac{1}{L^3}\sum_{q_i}^{q_{\rm max}}I(q_i)-\int\limits_{q<q_{\rm max}}\frac{d^3q}{(2\pi)^3} I(q)\Bigg],
\label{gtdim}
\end{align}
where
\ba
I(q)&=&\frac{2M_l}{2\omega_j(\vec q)\,\omega^\prime_j(\vec q)}
\frac{\omega_j(\vec q)+\omega^\prime_j(\vec q)}
{E^2-(\omega_j(\vec q)+\omega^\prime_j(\vec q))^2+i\epsilon},
\label{prop_contado}
\ea
and
\ba
\vec q&=&\frac{2\pi}{L}\,\vec n,
\vec{n} \in Z^3,
\ea
with $\omega_l(\vec q)=\sqrt{\vec{q}^2+m_l^2}$, $\omega^\prime_l=\sqrt{(\vec{q}^2+M_l^2)}$\cite{AlbertoKD}.

Instead of integrating over the energy states in the infinite volume, we sum over the discrete momenta allowed in a finite box of side $L$ with periodic boundary conditions. The three dimensional sum in Eq.~(\ref{gtdim}) can be reduced to one dimension considering the multiplicities of the cases having the same $\vec{n}^{\,2}$.

\begin{figure}
\includegraphics[width=0.65\textwidth]{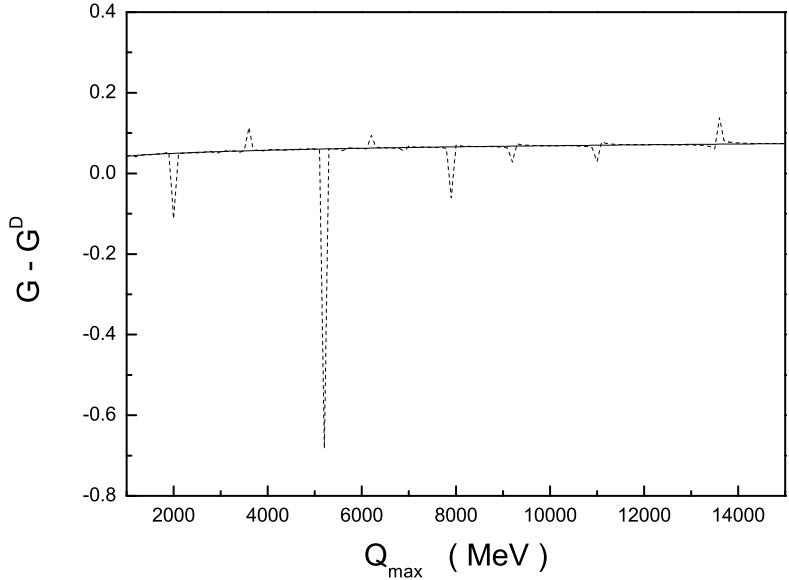}
\caption{Real part of the last two terms of the right hand side of Eq.~(\ref{gtdim}) for the $\bar{K} \Lambda$ channel. The solid line indicates the analytical form in Ref.~\cite{Olleroset}, while the dashed line represents oscillation results calculated with Eq.~(\ref{gtdim}) directly.}
\label{fig:oscilation}
\end{figure}

When calculating the limit of $q_{max}$ going to infinity in Eq.~(\ref{gtdim}) we obtain oscillations which gradually vanish as $q_{max}$ goes to infinity, as shown in Fig.~\ref{fig:oscilation}. Note that the imaginary part of $G^D$ and of the integral in Eq.~(\ref{gtdim}) are identical and they cancel in the construction of $\tilde G$, which is a real function.
The integral in Eq.~(\ref{gtdim}) has an analytical form as shown in the appendix part of Ref.~\cite{Olleroset}(See erratum), and it repeats the value calculated with Eq.~(\ref{gtdim}), but without fluctuation.

The eigenenergies of the box correspond to energies  that produce poles in the $T$ matrix.  Thus we search for these energies by looking for zeros of the determinant of $1-\tilde V\tilde G$

\be
\label{eq:det}
\det(1-\tilde V \tilde G)=1-\tilde V_{11}\tilde G_1-\tilde V_{22}\tilde G_2
+(\tilde V_{11} \tilde V_{22}- \tilde V_{12}^2)\tilde G_1\tilde G_2=0\, .
\ee

\begin{figure}
\includegraphics[width=0.65\textwidth]{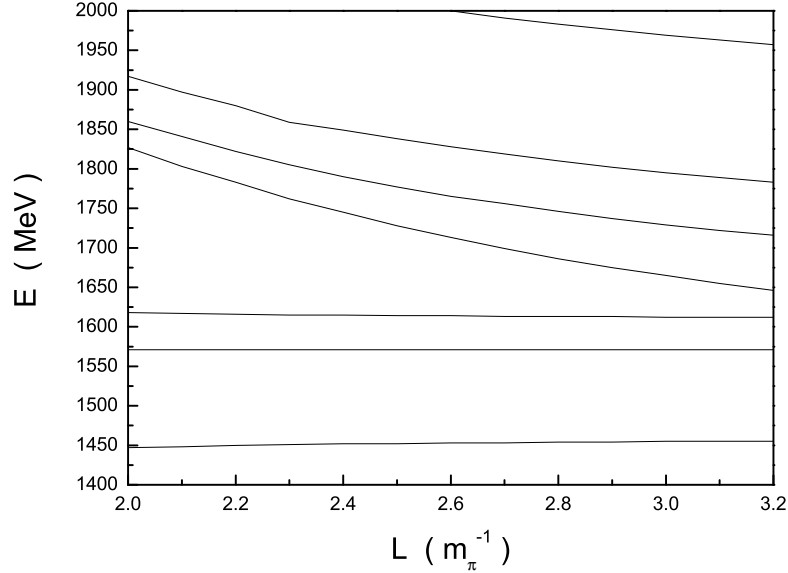}
\caption{Energy levels as functions of the cubic box size $L$, derived
from the chiral unitary approach in a box and using
$\bar{G}(E)$ from Eq.~(\ref{gtdim}).}
\label{fig:levelstwo}
\end{figure}

The energy levels obtained in the box for different values of $L$ are depicted in Fig.~\ref{fig:levelstwo}, and
a smooth behavior of energy levels as a function of $L$ is observed.
In Fig.~\ref{fig:levelstwo}, the first 3 energy levels are almost invariant when the cubic box size $L$ increases. Especially, the lowest and third levels are close to the $\pi \Xi$ and $\bar{K} \Lambda$ thresholds, respectively, therefore, they do not correspond to bound states of the pseudoscalar meson and baryon octet, but indicate the threshold effect in the finite volume.
The second level lies at 1570MeV, which is higher than the $\pi \Xi$ threshold, and can be regarded as a resonance state with strangeness $S=-2$ and isospin $I=1/2$. Apparently, this energy level is far away from the mass of the $\Xi(1620)$ particle announced by Belle collaboration.

\section{Summary}
\label{sect:summary}

In this work, the interaction of the pseudoscalar meson and the
baryon octet with strangeness $S=-2$ and isospin $I=1/2$ is investigated
by solving the Bethe-Salpeter equation in the unitary coupled-channel
approximation. It is found that a resonance state is generated dynamically
around 1550MeV, which owns a decay width about 120-200MeV. Thus this resonance state is not consistent with the $\Xi(1620)$ particle announced by Belle collaboration.
The coupling constants of this resonance state to different channels are calculated,
and it couples strongly to the $\pi \Xi$ channel.
Furthermore, this problem is also studied by solving the Bethe-Salpeter equation
in the finite volume, and the energy levels at different cubic box sizes are obtained.
It is found that the second energy level near 1570MeV might be a resonance state of the pseudoscalar meson and baryon octet, while the first and third levels might comes from the $\pi \Xi$ and $\bar{K} \Lambda$ thresholds, respectively.

\newpage

\end{document}